\documentclass[prd,twocolumn,aps,nofootinbib,nobibnotes,superscriptaddress,preprintnumbers]{revtex4-1}

\usepackage{mathrsfs}
\usepackage{yfonts}
\usepackage{tipa}
\usepackage{bm}
\usepackage{bbm}
\usepackage{amsmath}
\usepackage{amssymb}
\usepackage{amsthm}
\usepackage{amsfonts}
\usepackage{mathtools}
\usepackage{color}
\usepackage{latexsym}
\usepackage{relsize}
\usepackage[greek,english]{babel}
\usepackage{booktabs}
\usepackage[iso-8859-7]{inputenc}
\usepackage[dvipsnames]{xcolor}
\definecolor{refs}{RGB}{245,156,74}
 \usepackage[colorlinks=true,hyperfootnotes=true,citecolor=cyan]{hyperref}
\usepackage{array}
\newcolumntype{?}{!{\vrule width 2pt}}

\newcommand{\lp}{\left(}
\newcommand{\rp}{\right)}
\newcommand{\lb}{\left[}
\newcommand{\rb}{\right]}

\newcommand{\mR}{\mathcal{R}}

\newcommand{\mN}{\mathcal{N}}

\newcommand{\mL}{{\mathcal L}}
\newcommand{\mS}{{\mathcal S}}

\newcommand{\lsim}   {\mathrel{\mathop{\kern 0pt \rlap
  {\raise.2ex\hbox{$<$}}}
  \lower.9ex\hbox{\kern-.190em $\sim$}}}
\newcommand{\gsim}   {\mathrel{\mathop{\kern 0pt \rlap
  {\raise.2ex\hbox{$>$}}}
  \lower.9ex\hbox{\kern-.190em $\sim$}}}
\newcommand{\bw}{\begin{widetext}\begin{equation}}
\newcommand{\ew}{\end{equation}\end{widetext}}
\newcommand{\be}{\begin{equation}}
\newcommand{\ee}{\end{equation}}
\newcommand{\ba}{\begin{eqnarray}}
\newcommand{\ea}{\end{eqnarray}}

\newcommand{\diff}{{{\rm d}}}

\newcommand{\mH}{\mathcal{H}}

\newcommand{\nn}{\nonumber}

\newcommand{\Qr}{\mathring{Q}}
\newcommand{\zetah}{\hat{\zeta}}

\newcommand{\mpl}{M_{\rm Pl}}

\begin{document}

\title{Cosmology in $f(Q)$ geometry}

\author{Jose Beltr\'an Jim\'enez}
\email{jose.beltran@usal.es}
\address{Departamento de F\'isica Fundamental and IUFFyM, Universidad de Salamanca, E-37008 Salamanca, Spain.}

\author{Lavinia Heisenberg}
\email{lavinia.heisenberg@phys.ethz.ch}
\address{Institute for Theoretical Physics, 
ETH Zurich, Wolfgang-Pauli-Strasse 27, 8093, Zurich, Switzerland}

\author{Tomi Sebastian Koivisto}
\email{tomik@astro.uio.no}
\address{Nordita, KTH Royal Institute of Technology and Stockholm University, Roslagstullsbacken 23, SE-10691 Stockholm, Sweden}
\address{Laboratory of Theoretical Physics, Institute of Physics, University of Tartu, W. Ostwaldi 1, 50411 Tartu, Estonia}
\address{National Institute of Chemical Physics and Biophysics, R\"avala pst. 10, 10143 Tallinn, Estonia}

\author{Simon Pekar}
\email{simonale@uoguelph.ca}
\address{Institute for Theoretical Physics, 
ETH Zurich, Wolfgang-Pauli-Strasse 27, 8093, Zurich, Switzerland}

\date{\today}

\begin{abstract}
The universal character of the gravitational interaction provided by the equivalence principle motivates a geometrical description of gravity. The standard formulation of General Relativity \`a la Einstein attributes gravity to the spacetime curvature, to which we have grown accustomed. 
However, this perception has masked the fact that two alternative, though equivalent, formulations of General Relativity
in flat spacetimes exist, where gravity can be fully ascribed either to torsion or to non-metricity.
The latter allows a simpler geometrical formulation of General Relativity that is oblivious to the affine spacetime structure.
Generalisations along this line permit to generate teleparallel and symmetric teleparallel theories of
gravity with exceptional properties. In this work we explore modified gravity theories based on non-linear extensions of the non-metricity
scalar. After presenting some general properties and briefly studying some interesting background cosmologies (including accelerating solutions with relevance for inflation and dark energy), we analyse the behaviour of the cosmological perturbations. Tensor perturbations feature a re-scaling of the corresponding Newton's constant, while vector perturbations do not contribute in the absence of vector sources. In the scalar sector we find two additional propagating modes, hinting that $f(Q)$ theories introduce, at least, two additional degrees of freedom. These scalar modes disappear around maximally symmetric backgrounds because of the appearance of an accidental residual gauge symmetry corresponding to a restricted diffeomorphism. We finally discuss the potential strong coupling problems of these maximally symmetric backgrounds caused by the discontinuity in the number of propagating modes.
\end{abstract}

\preprint{NORDITA-2019-054}

\maketitle

\section{Introduction}

The gravitational phenomena can be interpreted as a manifestation of having a curved spacetime and this interpretation is possible thanks to the equivalence principle. 
The tremendous implication of this assumption is that the gravitational interaction is totally oblivious to the type of matter fields. However,
still within the geometrical framework, the curvature is not the unique geometrical object to represent the affine properties of a manifold \cite{schrodinger1985space,Olmo:2011uz,Heisenberg:2018vsk,BeltranJimenez:2019tjy}.
In fact, besides the curvature, the other two fundamental objects associated to the connection of a metric space are torsion and non-metricity.
In standard General Relativity (GR) \`a la Einstein, both non-metricity and torsion vanish. 

If we embrace the geometrical character of gravity advocated by the equivalence principle, it is pertinent to explore in which equivalent manners gravity can be geometrized. An equivalent representation of GR arises if one considers a flat spacetime with a metric but asymmetric connection. In this 
teleparallel description gravity is entirely assigned to torsion. A third equivalent and simpler representation of GR can be 
constructed on an equally flat spacetime without torsion, in which gravity is this time ascribed to non-metricity. Hence, the same underlying physical
theory, GR, can be described by the Einstein-Hilbert action $\int \sqrt{-g}R(g)$, the action of Teleparallel Equivalent of GR $\int \sqrt{-g}T$ \cite{Aldrovandi:2013wha} and Coincident GR $\int \sqrt{-g}Q$ \cite{BeltranJimenez:2017tkd} that rests on a symmetric teleparallel geometry \cite{Nester:1998mp}.
The geometrical trinity of GR and its concise manifestation in detail can be found in \cite{BeltranJimenez:2019tjy}. 

The fundamental basis of these geometrical interpretations offers a promising road for modified gravity. The equivalent descriptions of GR
with curvature, torsion and non-metricity represent different alternative starting points to modified gravity theories once the corresponding scalar quantities are for instance promoted 
to arbitrary functions thereof. Since modified gravity theories based on $f(R)$ \cite{Starobinsky:2007hu,Amendola:2010bk,Capozziello:2011et} and $f(T)$ \cite{Ferraro:2006jd,Ferraro:2011us,Hohmann:2019nat} are widely studied in the literature, we will focus on the less studied case of $f(Q)$ theories, which was introduced for the first time in \cite{BeltranJimenez:2017tkd}. Since at the cosmological background level, models based on $f(Q)$ are 
indistinguishable from $f(T)$ models, we will pay special attention to the study of perturbations and their distinctive properties in $f(Q)$ theories. Concerning perturbations on top of a FLRW 
(Friedmann-Lema{\^i}tre-Robertson-Walker)
background, models based on $f(T)$ seem to suffer from strong coupling problems because some of the genuine physical degrees of freedom lose their kinetic term at the quadratic order and, consequently, the standard 
perturbation theory breaks down on these backgrounds \cite{Golovnev:2018wbh}. 
At this stage, the number and nature of new propagating degrees of freedom in $f(T)$ models on an arbitrary background is still under investigation in the literature, see \cite{Ferraro:2018axk,Ferraro:2018tpu} and related studies \cite{Koivisto:2018loq,Blixt:2018znp,Blixt:2019ene}. 
 In this work, we will investigate whether models of modified gravity based on $f(Q)$ suffer from a similar strong coupling issue or whether it can be 
avoided for some interesting FLRW backgrounds relevant for dark energy and inflation.

Besides the $f(Q)$ models \cite{BeltranJimenez:2017tkd}, a few other classes of modified or extended gravity theories have been considered within the same symmetric teleparallel framework\footnote{Other recent studies into
torsion and non-metricity include e.g. \cite{Klemm:2018bil,Rasanen:2018ihz,Shimada:2018lnm,Krssak:2018ywd,Janssen:2018exh,Iosifidis:2018jwu,Penas:2018mnc,Iosifidis:2019jgi,Janssen:2019doc}.}. The so called 
Newer GR \cite{BeltranJimenez:2017tkd,BeltranJimenez:2018vdo,Hohmann:2018wxu,Soudi:2018dhv} is a five-parameter quadratic extension of the action $\int \sqrt{-g}Q$, which has been also considered in the language of 
differential forms \cite{Adak:2005cd,Adak:2008gd,Koivisto:2018aip} and extended to arbitrary derivative order by including, in the most general case, ten free functions of the d'Alembertian operator \cite{Conroy:2017yln}.
Parity-odd quadratic terms were taken into account in an exhaustive analysis of scale-invariant metric-affine theories \cite{Iosifidis:2018zwo}. Without the restriction to quadratic terms, cosmological solutions based on Noether symmetries were derived for generic first derivative order non-metric actions \cite{Dialektopoulos:2019mtr}.
J{\"a}rv {\it et al} introduced a non-minimal coupling of a scalar field to the $Q$-invariant \cite{Jarv:2018bgs}, which then R{\"u}nkla and Vilson generalised by coupling the scalar to
the five Newer GR terms \cite{Runkla:2018xrv}. Nonminimal couplings to matter were also considered in extended $f(Q)$ models by Harko {\it et al} \cite{Harko:2018gxr,Lobo:2019xwp}. 
To probe the potential 
cosmological viability of these and other classes of models, it is natural to begin by uncovering the propagating degrees of freedom of the prototype $f(Q)$ models in the FLRW background. 

This paper is structured as follows. After reviewing the theory in Section \ref{theory}, we briefly look at the FLRW background equations in Section \ref{background}. The main analysis of the perturbations and
the structure formation is found in Section \ref{perturbations}. To provide a better understanding and a more clear interpretation of the perturbations we discuss the gauge transformation properties of the $Q$-scalar in Section \ref{gauge}. Section \ref{conclusions} presents a concise summary of our findings and points out the direction for further research that needs to be carried in order to fully uncover the cosmological potential of $f(Q)$ gravity. 

\section{The Theory}
\label{theory}

We will use the so-called Palatini formalism where the metric and the connection are treated on equal footing so they are independent objects whose relation is only imposed by the field equations. In this framework, our spacetime manifold is endowed with a metric structure determined by the metric $g_{\mu\nu}$, while the affine connection $\Gamma^\alpha{}_{\mu\nu}$ provides the affine structure that stipulates how tensors are transported, thus defining the covariant derivative. In the class of theories under consideration in this work, the fundamental object is the non-metricity tensor defined as 
$Q_{\alpha\mu\nu}=\nabla_\alpha g_{\mu\nu}$. It manifests the failure of the connection in being metric compatible.
From the non-metricity tensor $Q_{\alpha\mu\nu}$, we can derive the disformation
\be
L^\alpha{}_{\mu\nu} = \frac{1}{2}Q^\alpha{}_{\mu\nu} - Q_{(\mu\nu)}{}^\alpha\,,
\ee 
that measures how far from the Levi-Civita connection the symmetric part of the full connection is. It will also be convenient to introduce the non-metricity conjugate defined as
\begin{equation}
P^\alpha{}_{\mu\nu} = -\frac{1}{2}L^\alpha{}_{\mu\nu} + \frac{1}{4}\lp Q^\alpha - \tilde{Q}^\alpha\rp g_{\mu\nu} - 
\frac{1}{4}\delta^\alpha_{(\mu}Q_{\nu)}\,,
\end{equation}
where the two independent traces $Q_\alpha=g^{\mu\nu}Q_{\alpha\mu\nu}$ and $\tilde{Q}_\alpha=g^{\mu\nu}Q_{\mu\alpha\nu}$ of the non-metricity tensor enter. Let us now introduce the non-metricity scalar that will play a central role in this work
\be
Q = -Q_{\alpha\mu\nu}P^{\alpha\mu\nu}\,.
\ee
One can then see why we call $P^{\alpha\mu\nu}$ the non-metricity conjugate because it satisfies
\be
P^{\alpha\mu\nu}=-\frac12\frac{\partial Q}{\partial Q_{\alpha\mu\nu}}.
\ee
We can then use the non-metricity scalar to write the action for the theories that we consider as
\begin{equation}  \label{action}
\mS = \int {\mathrm{d}}^4 x \sqrt{-g}\left[ {-}\frac{1}{2}f(Q) + \mL_M\right]\,,
\end{equation}
where $\mL_M$ stands for the matter Lagrangian. The motivation for the particular choice of the non-metricity scalar and the above action is that GR is reproduced (classically, up to a boundary term) for the choice $f=Q/8\pi G$ \cite{BeltranJimenez:2017tkd}, i.e., for this choice we recover the so-called Symmetric Teleparallel Equivalent of GR.

Before proceeding any further, it will be convenient to clarify that the geometrical framework that we use has a flat and torsion-free connection so that it must correspond to a pure coordinate transformation from the trivial connection as explained in \cite{BeltranJimenez:2017tkd}. More explicitly, the connection can be parameterised with a set of functions $\xi^\alpha$ as\footnote{It must be understood in \eqref{Connectionxi} that $\xi^\alpha=\xi^\alpha(x^\mu)$ is an invertible relation and $\frac{\partial x^\alpha}{\partial \xi^\rho}$ is the inverse of the corresponding Jacobian.}
\be
\Gamma^\alpha{}_{\mu\beta}=\frac{\partial x^\alpha}{\partial \xi^\rho}\partial_\mu\partial_\beta\xi^\rho\,.
\label{Connectionxi}
\ee
Therefore, it is always possible to make a coordinate choice so that the connection vanishes (more specifically, any coordinates affinely related to $\xi^\alpha=x^\alpha$.). We call these coordinates the coincident gauge and we will denote quantities evaluated in this gauge with a ring over the corresponding symbol so, by definition $\mathring{\Gamma}^\alpha{}_{\mu\nu}=0$. Thus, in the coincident gauge we will have
\be
\Qr_{\alpha\mu\nu}=\partial_\alpha g_{\mu\nu},
\ee
while in an arbitrary gauge we have
\ba
Q_{\alpha\mu\nu}&=&\partial_\alpha g_{\mu\nu}-2\Gamma^\lambda{}_{\alpha(\mu}g_{\nu)\lambda}\nonumber\\
&=&\Qr_{\alpha\mu\nu}-2\frac{\partial x^\lambda}{\partial \xi^\rho}\partial_\alpha\partial_{(\mu}\xi^\rho g_{\nu)\lambda}.
\ea

The metric field equations can be written as
\ba
&&\frac{2}{\sqrt{-g}}\nabla_\alpha\left(\sqrt{-g}f_QP^\alpha{}_{\mu\nu}\right)
+ \frac{1}{2}g_{\mu\nu} f \nonumber\\
&&+ f_Q\left( P_{\mu\alpha\beta}Q_{\nu}{}^{\alpha\beta}
-2Q_{\alpha\beta\mu}P^{\alpha\beta}{}_\nu\right) = T_{\mu\nu} \,,\label{efe}
\ea
where $f_{Q}=\partial f/\partial Q$. By raising one index this adopts an even slightly more compact form,
\be
\frac{2}{\sqrt{-g}} \nabla_\alpha\left(\sqrt{-g}f_{Q}P^{\alpha\mu}{}_\nu\right) 
 +  \frac{1}{2}\delta^\mu_\nu f 
 +  f_{Q} P^{\mu\alpha\beta}Q_{\nu\alpha\beta}   =   T^{\mu}{}_{\nu} \,.\label{efe2}
\ee
{The connection equation of motion can be straightforwardly computed by noticing that the variation of the connection with respect to $\xi^\alpha$ is equivalent to performing a diffeomorphism so that $\delta_\xi\Gamma^\alpha{}_{\mu\beta}=-\mathcal{L}_\xi \Gamma^\alpha{}_{\mu\beta}=-\nabla_\mu\nabla_\beta\xi^\alpha$, where we have used that the connection is flat and torsion-free.} Thus, in the absence of hypermomentum the connection field equations read
\begin{equation}
\nabla_\mu\nabla_\nu \left(\sqrt{-g} f_Q P^{\mu\nu}{}_\alpha\right) = 0\,.  \label{cfe}
\end{equation}
From the metric and the connection equations one can verify that $\mathcal{D}_\mu T^\mu{}_\nu=0$, where $\mathcal{D}_\mu$ is the metric-covariant derivative \cite{BeltranJimenez:2018vdo}, as it should by virtue of diffeomorphism (Diff) invariance. In the most general case with a non-trivial hypermomentum, one would obtain a relation between the divergence of the energy-momentum tensor and the hypermomentum \cite{Harko:2018gxr}. \\

In order to clarify the conservation in the matter sector and the Bianchi identities, let us look at the gauge invariance of the theory. Under a Diff  given by $x^\mu\rightarrow x^\mu+\zeta^\mu$, the non-metricity scalar changes as
\be \label{change}
\delta_\zeta Q  =  \frac{\partial Q}{\partial g_{\mu\nu,\alpha}}\delta_\zeta g_{\mu\nu,\alpha} + \frac{\partial Q}{\partial g_{\mu\nu}}\delta_\zeta g_{\mu\nu} + 
\frac{\partial Q}{\partial \Gamma^\alpha{}_{\mu\nu}}\delta_\zeta \Gamma^\alpha{}_{\mu\nu}\,.
\ee 
{The first term is 
\be
 \frac{\partial Q}{\partial g_{\mu\nu,\alpha}}\delta_\zeta g_{\mu\nu,\alpha}  = 2P^{\alpha\mu\nu}\mathcal{L}_\zeta \mathring{Q}_{\alpha\mu\nu}\,,
 \ee
where 
\begin{align}
\mL_\zeta\Qr_{\alpha\mu\nu}=&\;\zeta^\rho\partial_\rho\partial_\alpha g_{\mu\nu}+\partial_\alpha\zeta^\rho \partial_\rho g_{\mu\nu}+2\partial_\alpha g_{\rho(\mu}\partial_{\nu)}\zeta^\rho\nonumber\\&+2g_{\rho(\mu}\partial_{\nu)}\partial_\alpha\zeta^\rho\,.
\end{align}
The last term in (\ref{change}) is
\ba
&&\frac{\partial Q}{\partial \Gamma^\alpha{}_{\mu\nu}}\delta_\zeta \Gamma^\alpha{}_{\mu\nu} = \frac{\partial Q}{\partial Q_{\alpha\beta\gamma}}\frac{\partial Q_{\alpha\beta\gamma}}{\partial \Gamma^\kappa{}_{\mu\nu}}\mathcal{L}_\zeta \Gamma^\kappa{}_{\mu\nu} \nonumber\\
&&= -2P^{\alpha\beta\gamma}\lp-2\delta^\mu_\alpha \delta^\nu_{(\beta} g_{\gamma)\lambda}\rp (-\nabla_\mu\nabla_\nu\zeta^\lambda )\nonumber\\
&&= -4P^{\alpha\beta\gamma}g_{\lambda(\beta}\nabla_{\gamma)}\nabla_\alpha\zeta^\lambda\,,
\ea
It is possible to show that the non-metricity scalar varies under Diff as
\be
\delta_\zeta Q  =  \frac{\partial Q}{\partial Q_{\alpha\mu\nu}}\mathcal{L}_\zeta Q_{\alpha\mu\nu} +  \frac{\partial Q}{\partial g_{\mu\nu}}\mathcal{L}_\zeta g_{\mu\nu} = -\mL_\zeta Q\,,
\ee
and thus the action (\ref{action}) is Diff-invariant. However, when fixed to the coincident gauge, the action no longer remains generally Diff-invariant. One obtains
\ba
\delta_\zeta \mathring{\mS} &=& 
 -2\int \diff^4 x\zeta^\lambda\Big[ \sqrt{-g}\lp\partial_\alpha\partial_\gamma f_Q\rp \mathring{P}^{\alpha\gamma}{}_\lambda \nonumber \\
 & + & 2\lp\partial_{(\alpha}f_Q\rp\lp \partial_{\gamma)}\sqrt{-g}\mathring{P}^{\alpha\gamma}{}_\lambda\rp\Big]\,.
 \ea
We have dropped a total derivative and used the identity
\be
\partial_\alpha \partial_\mu (\sqrt{-g}\mathring{P}^{\alpha\mu}{}_\nu)=0\,.
\ee 
Thus, only when $f_{QQ}=0$, the action is in general Diff invariant. However, for a generic $f$ there appears a term which vanishes on-shell as the result of the non-trivial extra set of equations (\ref{cfe}).}

\section{Cosmological background evolution}
\label{background}

As we mentioned above, at the cosmological background level our model based on \eqref{action} does not differ from the $f(T)$ models.
It also shares similar cosmological solutions as the vector distortion parametrisation \cite{Jimenez:2015fva,Jimenez:2016opp}.
The background evolution and the different self-accelerating solutions were briefly discussed in \cite{BeltranJimenez:2017tkd}. We shall only quickly 
review the background equations of motion and their defining properties. {We will consider the FLRW metric
\be \label{flrw}
\diff s^2=-N^2(t)\diff t^2+a^2(t)\diff\vec{x}^2\,,
\ee
where $N(t)$ and $a(t)$ are the lapse function and the scale factor, respectively. From now on and unless otherwise stated, we will fix the coincident gauge so the connection is trivial. The non-metricity scalar is then $Q=6\frac{H^2}{N^2}$. Since we have used diffeomorphisms to fix the coincident gauge, one could think that we are not allowed to select any particular lapse function. However, the special case of $f(Q)$ theories does allow so because $Q$ retains a residual time-reparameterisation invariance, as already explained in \cite{BeltranJimenez:2017tkd} so we will use this symmetry to set $N=1$.}

The cosmological 
equations of motion are given by 
\begin{eqnarray}\label{BGeqsfQ}
6f_QH^2-\frac12f&=&\rho \label{eqFrid}, \nonumber\\
\big(12H^2f_{QQ}+f_Q\big)\dot{H}&=&-\frac{1}{2}(\rho+p)\,.
\end{eqnarray}
The standard matter fields satisfy the continuity equation $\dot{\rho}=-3H(\rho+p)$, that is consistent with the above cosmological equations thanks to the time-reparameterisation invariance aforementioned. These equations are formally the same as those of the $f(T)$ theories so we refer to \cite{Hohmann:2017jao} for an extensive analysis of the background cosmology based on those equations. We will content ourselves with discussing some interesting features of the background cosmology for these theories.

There is a particularly interesting class of theories that give a background evolution identical to that of General Relativity. Such a class of models can be easily obtained by imposing $Qf_Q-\frac12f=\frac{Q}{16\pi G}$ whose solution is 
\be
f=\frac{1}{8\pi G}\Big(Q+M\sqrt{Q}\Big)
\ee
with $M$ some mass scale. Of course, $M=0$ corresponds to the GR equivalent, but it is remarkable that there exists a whole class of theories whose background cosmology is the same as in GR for any matter content. The different values of $M$ could only be discriminated by analysing the evolution of the perturbations and this property by itself makes this particular choice of $f(Q)$ an interesting study case since the evolution of the perturbations can be modified while maintaining the background oblivious to such modifications.

On the other hand, the STEGR supplemented with a general power-law term, i.e. 
\be
f=\frac{1}{8\pi G}\left[Q-6\lambda M^2\left(\frac{Q}{6M^2}\right)^\alpha\right]
\ee
with $\lambda$ and $\alpha$ dimensionless parameters, gives rise to branches of solutions applicable either to early universe cosmology or to dark energy depending on the value of $\alpha$. For these models, the Friedmann equation modifies into  
\be
H^2\left[1+(1-2\alpha)\lambda\left(\frac{H^2}{M^2}\right)^{\alpha-1}\right]=\frac{8\pi G}{3}\rho\,.
 \ee
Again, for $\alpha=1/2$ we recover the aforementioned class of theories with the same background evolution as GR, while $\alpha=1$ is of course degenerate with the STEGR and can be fully absorbed into $G$. It is then apparent from the above modified Friedmann equation that the corrections to the usual GR evolution will appear at low curvatures for $\alpha<1$, while for $\alpha>1$, the corrections will be relevant in the high curvatures regime. Thus, theories with $\alpha>1$ will be relevant for the early universe (with potential applications to inflationary solutions),  whereas theories with $\alpha<1$ will give corrections to the late-time cosmology, where they can give rise to dark energy.

We would like to bring the equations into an autonomous-like form. In order to do that we introduce the dimensionless variables
\begin{equation}
x_1=\frac{\rho_m}{Qf_Q}, \quad x_2=\frac{\rho_r}{Qf_Q} \quad \text{and} \quad x_3=\frac{f}{2Qf_Q}\,.
\end{equation}
We have assumed that the total energy density $\rho$ is comprised by a matter $\rho_m$ and a radiation $\rho_r$ component. In terms of the variables the dynamical system takes the form
\begin{eqnarray}
&&x_1'=x_1(\epsilon -3)+3x_1^2+4x_1x_2\nonumber\\
&&x_2'=x_2(\epsilon -4)+3x_1x_2+4x_2^2\nonumber\\
&&x_3'=\epsilon(x_3-1)+3x_1x_3+4x_2x_3\,,
\end{eqnarray}
where $\epsilon=-\dot{H}/H^2$ and prime denotes the derivative with respect to {$\log{a}$. The independent variables can be taken as $\{x_1,x_2,\epsilon\}$. The Friedmann equation \eqref{eqFrid} imposes the constraint equation}
\begin{equation}
x_1+x_2+x_3=1\,.
\end{equation}
One can easily solve the system for its critical points. We find the following three critical points of the dynamical system
\begin{eqnarray}
&&{\rm I}:\quad \; \left( x_1=0, \quad x_2=0, \quad x_3=1\right)\nonumber\\
&&{\rm II}: \;\;\left( x_1=1-\frac13\epsilon, \quad x_2=0, \quad x_3=\frac13\epsilon\right)\nonumber\\
&&{\rm III}: \left( x_1=0, \quad x_2=1-\frac14\epsilon, \quad x_3=\frac14\epsilon\right)\,.
\end{eqnarray}
We can include small perturbations and consider the following linearised system
\begin{equation}
\frac{d}{dt}
\begin{pmatrix}
\delta x_1\\
\delta x_2\\
\delta x_3 
\end{pmatrix}=M\begin{pmatrix}
\delta x_1\\
\delta x_2\\
\delta x_3 
\end{pmatrix}
\end{equation}
with the matrix $M$ of the dynamical system given by $M=$
\begin{equation}
\begin{pmatrix}
6x_1+4x_2+\epsilon-3 & 3x_2 & 3x_3 \\
4x_1 & 3x_1+8x_2+\epsilon-4 & 4x_3  \\
0 & 0 & 3x_1+4x_2+\epsilon
\end{pmatrix}\,.
\end{equation}
The stability analysis of the critical points reveals that the first critical point $I$ is a stable attractor, the second one $II$ is a saddle point and the third one $III$ is an unstable repeller.

In the following we will study two explicit examples with some interesting properties.

\begin{figure*}[ht!]
   \centering
 \includegraphics[width=8.9cm]{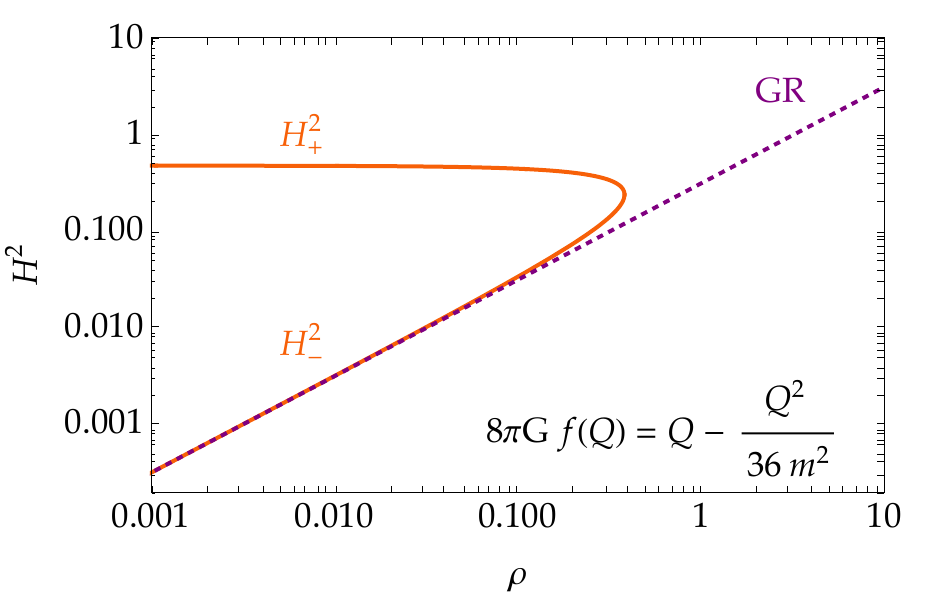}
 \includegraphics[width=8.9cm]{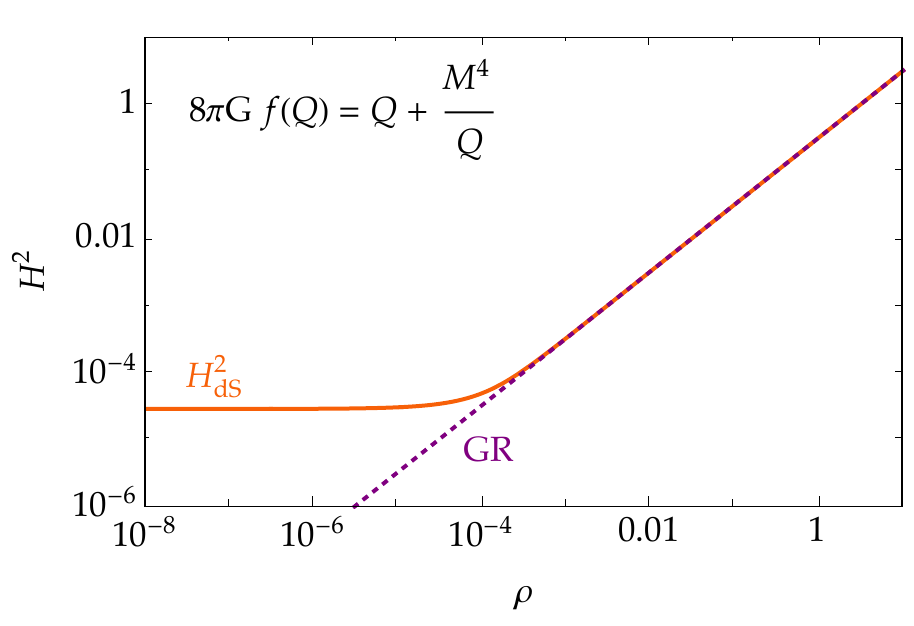}
\caption{In these figures we show the Hubble function $H^2$ as a function of $\rho$ for the two examples discussed in the main text. We have normalised to the Planck mass and fixed the parameters to $m=0.5\mpl$  in the left panel and $M=10^{-2}\mpl$ in the right panel.}
\label{H2examples}
\end{figure*}

\textbf{Example 1: $8\pi Gf(Q)= Q-\frac{1}{36}(Q/m)^2$}\\
In this case, the Friedmann equation in \eqref{BGeqsfQ} simplifies to 
\be
3H^2\left(1-\frac{H^2}{2m^2}\right)=8\pi G\rho
\ee
which has two branches of cosmological solutions 
\be
H^2_\pm=m^2\left(1\pm\sqrt{1-\frac{16\pi G\rho}{3m^2}}\,\right)\,. 
\label{eqH2example1}
\ee
The cosmological background evolution for this example is formally the same as the one obtained in \cite{Jimenez:2016opp} from a completely different framework based on a  generalised Weyl geometry with vector distortion giving both non-metricity and torsion, first introduced in \cite{Jimenez:2015fva}. Thus, we will limit ourselves to summarise the main properties of these solutions and we will refer to \cite{Jimenez:2016opp} for a more detailed analysis. The most remarkable feature of this cosmology is the existence of a maximum allowed density in the universe given by $8\pi G\rho_{\rm max}=\frac32 m^2$, which is enforced by the square root in \eqref{eqH2example1}\footnote{The existence of a maximum energy density enforced by some square root structure is the main motivation behind the Born-Infeld-inspired theories of gravity (see \cite{BeltranJimenez:2017doy} and references therein).}. Below the maximum density, we have two branches: $H_-^2$ that approaches the usual GR evolution at low densities and $H_+^2$ that gives an approximate de Sitter universe with $H^2_{\rm dS}\simeq 2m^2$ regardless the value of $\rho$ (see left panel in Fig. \ref{H2examples}).

\textbf{Example 2: $8\pi Gf(Q)=Q+M^4/Q$}\\
The Friedmann equation in this case becomes 
\be
H^2\left(1-\frac{M^4}{12H^4}\right)=8\pi G\rho
\ee
that can be solved for $H^2$ as
\be
H^2_\pm=\frac{4\pi G}{3}\rho\left(1\pm\sqrt{1+\frac{3M^4}{64\pi^2 G^2\rho^2}}\right).
\ee
The negative branch is not physical because it gives $H^2_-<0$, so only the branch $H^2_+$ is physical. As in the previous example, this evolution was also obtained in the framework of generalised Weyl geometries in \cite{Jimenez:2016opp}. The physical branch $H_+^2$ recovers the usual GR Friedmann equation for $\rho\gg M^2 \mpl^2$, while at low densities the Hubble function approaches the constant $H_{\rm dS}^2=\frac{M^2}{2\sqrt{3}}$ corresponding to an asymptotically de Sitter solution, as shown in the right panel of Fig. \ref{H2examples}. This example can thus be relevant for dark energy applications since it allows for solutions that naturally give a transition from a matter dominated universe to an accelerating de Sitter universe. However, as we will show in our analysis of the perturbations in the subsequent sections, this asymptotically de Sitter solutions are prone to strong coupling problems in the scalar sector of the perturbations, thus casting doubts on its phenomenological viability.

Since the $f(Q)$ theories share the background equations with the $f(T)$ theories, we will not investigate further the possible background cosmologies and we will now turn to the main focus of this paper, i.e., the evolution of the perturbations. Let us finish this brief discussion of the background cosmological evolution by advancing the existence of strong coupling problems at the perturbative level that may cast serious doubts on the viability of the background cosmologies.

\section{Cosmological perturbations}
\label{perturbations}

For the sake of generality, we will study the evolution of the cosmological perturbations in the presence of a $K$-essence field so our action is
\begin{equation}
\mathcal{S}_M=\int \diff^4x \sqrt{-g}\left(-\frac12 f(Q)+P(X)\right)\,,
\end{equation}
where $X=-\partial_\mu\chi\partial^\mu\chi$. The background Friedmann equations now read
\begin{align}
&f+P+12H^2f_Q-2P_X\dot{\bar{\chi}}^2=0\\
& P_X\dot{\bar{\chi}}^2+2\dot{H}\Big(f_Q-12H^2f_{QQ}\Big)=0\,.
\end{align}
On the other hand, the scalar field equation reads
\begin{equation}
\frac{\diff}{\diff t}\left(a^3P_X\,\dot{\bar{\chi}}\right)=0
\end{equation}
Since the crucial difference to $f(T)$ theories will arise at the level of perturbations, we will investigate in detail the second order action of cosmological perturbations.

For this purpose, we decompose the metric in terms of the irreducible representations of the background $SO(3)$ symmetry expressed in conformal time as follows
\cite{Mukhanov:1990me}
\begin{align}\label{perturbed_metric}
\delta g_{00} &= -2a^2\,\,\phi\,,\nonumber\\
\delta g_{0i} &={\delta g_{i0} = \,a^2\,\left(\partial_i B+B_i\right)\,,}\nonumber\\
\delta g_{ij} &= 2a^2 \left[\,{-}\varphi\delta_{ij} +\left(\partial_i\partial_j-\frac{\delta_{ij}}{3}\partial^k\partial_k\right)E+\partial_{(i}E_{j)}+h_{ij}\right]\,,
\end{align}
with the scalar perturbations $\phi$, $B$, $\varphi$ and $E$, the vector perturbations $B_i$ and $E_i$ satisfying $\partial^iB_i=0=\partial^iE_i$ and the tensor perturbations $h_{ij}$ with the properties $\partial^ih_{ij}=0=h_i{}^i$.
Similarly, we decompose the $K$-essence field into its background contribution and perturbation 
\begin{equation}
\chi(\tau,x,y,z)=\bar{\chi}(\tau)+\delta\chi(t,x,y,z).
\end{equation} 
As usual, the different sectors will decouple thanks to the background symmetries so we can treat them separately. We shall use conformal time, corresponding to the choice of lapse $N=a$ in (\ref{flrw}), by $\tau$. Derivatives with respect to the conformal time will be denoted by primes, and thus 
the conformal Hubble rate is defined as $\mathcal{H}=a'/a=aH$. A word on the choice of time variable is in order before proceeding. As we have discussed, the background equations retain time re-parameterisation invariance, so the choice of time coordinate is irrelevant there. However, since we are working in the coincident gauge, the perturbations do not enjoy Diffs invariance and the choice of background time coordinate might lead to different results. We have checked that all our conclusions below remain valid regardless the choice of the lapse function.

\subsection{Tensor perturbations}\label{tensormodes}
We introduce the tensor perturbations as the transverse traceless part of the metric fluctuations that can be decomposed into its two helicity modes $h_{(\lambda)}$. After decomposing the tensor field in Fourier modes with respect to the spatial coordinates and using the background equations of motion, the second order action becomes
\begin{equation}\label{action_tensormodes}
\mathcal{S}^{(2)}_{\rm tensor} = \frac{1}{2}\sum_\lambda\int \diff^3k\,\diff \tau\,\,a^2\, f_Q \left[(h'_{(\lambda)})^2-k^2h_{(\lambda)}^2  \right]\,.
\end{equation}
As it becomes clear from the above expression, the tensor perturbations are massless and have the same propagation speed of gravitational waves as in GR \cite{BeltranJimenez:2017tkd}. {This was to be expected from the form of the action for the theories under consideration because all the modifications to the pure tensor perturbations come from the term $Q_{\alpha\mu\nu} Q^{\alpha\mu\nu}\sim \partial_\alpha h_{ij}\partial^\alpha h^{ij}$. All the other independent scalars constructed out of the non-metricity vanish for the quadratic action. It is clear that tensor perturbations cannot contribute to the vector traces of the non-metricity at first order. The remaining independent scalar is $Q_{\alpha\mu\nu} Q^{\mu\alpha\nu}\sim \partial_{i}h_{jk} \partial^j h^{ik}$ which gives zero in the quadratic action via integration by parts and recalling that the background is homogeneous. Thus, our result is the expected one that the only modification to tensor perturbations is the appearance of the time-depending redressing of the effective Planck mass with $f_Q$. This has some observational consequences because the coupling constant of GWs to matter sources will be $G/f_Q$. As we will see, this is the same effective Newton's constant driving the growth of structures. On the other hand, the modified coupling constant will also affect the cosmological propagation of gravitational waves because the usual Hubble friction term acquires a $f_Q$-dependent correction. More explicitly, the propagation of GWs is governed by the modified equation
\be
h''_{(\lambda)}+2\mH\left(1+\frac{\diff \log f_Q}{2\mH\diff \eta}\right) h'_{(\lambda)}+k^2h_{(\lambda)}=0.
\ee
An immediate consequence is thus a modification on the luminosity-distance as measured from GWs. Due to the additional friction introduced by the time-dependent value of $f_Q$, the amplitude of GWs decays as $\vert h_{(\lambda)}\vert\propto 1/(a\sqrt{f_Q})$ instead of the usual $1/a$ dilution. We can then obtain constraints on $f_Q$ by comparing the luminosity-distance as measured with GWs and the one inferred from observing photons, assuming that their propagation is described by Maxwellian electromagnetism. A perfect candidate for this is the merger of two neutron stars as detected by the LIGO collaboration where the signals in GWs and the electromagnetic counterpart were measured. This modification on the luminosity-distance is formally analogous to what happens in models with extra-dimensions where gravity can leak into the additional dimensions \cite{Deffayet:2007kf}. The effect on the luminosity-distance arising from a time-variation of the effective Planck mass has also been analysed in \cite{Amendola:2017ovw} (see also \cite{Nishizawa:2017nef,Belgacem:2017ihm,Ezquiaga:2018btd} for the effects on the propagation of GWs arising from a modified friction term).}

\subsection{Vector perturbations}
Naively counted there are four vector modes in $E_i$ and $B_i$. However, one immediate observation is that the fields $B_i$ appear in the second order action as non-dynamical fields and they can be integrated out by means of an algebraic equation. We use their equations of motion in order to remove them from the second order action. Then, the remaining action solely depends on $E_i$. Nevertheless, once the background equations of motion are used, their contributions trivialize in the sense that the entire second order action vanishes. This can also be straightforwardly understood from the field equations directly, which can be written as
\begin{eqnarray}
k^2f_Q\left(\vec{B}-\vec{E}'\right)&=&0,\\
k^2\left[f_Q\left(\vec{B}-\vec{E}'\right)\right]'&=&0,
\label{eq:vecpert}
\end{eqnarray}
where we have made used of the background equations. We then see that the first equation is not dynamical and simply fixes $\vec{B}$ in terms of $\vec{E}'$. The second equation, being the derivative of the constraint, does not give new information.
This means that vector perturbations do not propagate, which coincides with the usual result in GR in the absence of vector sources. However, while in GR this is a consequence of diffeomorphisms invariance and the vector modes can be associated to the Lagrange multipliers of transverse Diffs, in the $f(Q)$ theories, once we fix the coincident gauge as we have done here, there is not, in principle, any symmetry ensuring that the vector perturbations are Lagrange multipliers and, consequently, one might expect that they become dynamical in the full theory. There is in fact a simple argument that clearly shows that at most two vector modes will become dynamical in the full theory, while the other two vector modes will remain non-propagating dof's. The reason is that, in the coincident gauge, the scalar $Q$ reduces to the Ricci scalar deprived of the total derivative. If we perform an ADM splitting where the metric is decomposed into the lapse function $\mathcal{N}$, the shift vector $\mathcal{N}_i$ and the spatial metric $\gamma_{ij}$, the non-metricity scalar $Q$ will only contain time derivatives of $\gamma_{ij}$ (see Appendix \ref{ADM} for more details) so that the $f(Q)$ action in the ADM formalism will have the form 
\be
\mS_{\rm ADM}=-\frac12\int\diff^3x\diff t\mathcal{N}\sqrt{-\gamma}f(\dot{\gamma}_{ij},\gamma_{ij},\mathcal{N},\mathcal{N}_i)
\label{eq:ADM}
\ee
where we clearly see the non-dynamical character of the lapse and the shift. This means that only the vector modes contained within $\gamma_{ij}$ can become dynamical, while those contributing to the shift remain non-dynamical at the full non-linear level. Thus, the non-dynamical nature of $\vec{B}$ reflects that the shift is not a propagating dof in the full theory. As for $\vec{E}$, it remains to see if it also corresponds to a non-dynamical field in the full theory. If this was not the case, $\vec{E}$ would represent a mode that becomes strongly coupled in the highly symmetric FLRW background.

 \subsection{Scalar perturbations}
There are four scalar modes $\{B, \phi, \varphi, E\}$ in the metric fluctuations and one in the fluctuation of the matter field $\delta\chi$. However, the two scalar perturbations $B$ and $\phi$ are not dynamical and can be immediately integrated out using their algebraic equations of motion. The non-dynamical nature of these modes can be understood from our previous discussion on the ADM decomposition because $B$ and $\phi$ are just the linearised lapse function and the longitudinal part of the perturbed shift, which are in fact non-dynamical in the full theory.
The remaining second order action depends on the scalar perturbations $\{\varphi, E, \delta\chi\}$. Its cumbersome form will not be necessary for us here so we will omit it\footnote{The full Hessian is reported in Eq. (\ref{fullhessian}). We give the corresponding equations of motion in Sec. \ref{Sec:equations}. }. It is interesting however to give the determinant of the corresponding Hessian, which is 
\be \label{hessiandet}
\frac{12 k^4 Q^2 f_Q f^2_{QQ}(f_Q+2Qf_{QQ})\epsilon^2}{12 f_Q^2+2Qf_Q f_{QQ}(9-5\epsilon)-Q^2f_{QQ}^2(8-3\epsilon)\epsilon}\,,
\ee
where we have introduced $\epsilon=1-\mH'/\mH^2$, which is the usual slow-roll parameter expressed in conformal time (though we do not assume it to be small here). The vanishing of the above expression indicates a degenerate system so that at least one of the remaining three scalar modes becomes non-dynamical. This is of course the case when $f_{QQ}=0$ as it corresponds to GR (we know that there will actually be two more non-dynamical modes in that case owed to Diffs-invariance). Since the determinant is proportional to $k^4$, we also see that the system becomes degenerate for homogeneous perturbations, which is a reflection of the additional time reparameterisation symmetry of the background. Besides these two limiting cases, there is another situation where the Hessian is degenerate, which corresponds to background solutions with $\epsilon=0$. These solutions describe maximally symmetric backgrounds, i.e.,  Minkowski and (anti-)de Sitter. The Hessian for these backgrounds reduces to the simple form
\[
\left(
\begin{array}{ccc}
 0 &0   &0   \\
  0&-\frac{2k^4 Q f_Q f_{QQ}}{6f_Q+9Qf_{QQ}}   &0   \\
  0& 0  & 1  
\end{array}
\right)
\]
that clearly shows how one of the scalar modes loses its kinetic term and becomes non-dynamical. Furthermore, it is not difficult to show that, after integrating out this new non-dynamical mode, the equation for the remaining scalar mode of the gravitational sector trivialises so that the full scalar sector disappears (assuming that the matter sector is simply given by an exact cosmological constant). More explicitly, the equations for $\varphi$ and $E$ in the presence of a cosmological constant and after integrating out the non-dynamical modes $\phi$ and $B$ are given by
\begin{align}
&\frac{k^4 Q f_Q f_{QQ}}{H^2(2f_Q+3 Q f_Q)}\Big(3H E'+k^2 E-3\varphi\Big)=0,\\
&\frac{k^4 Q f_Q f_{QQ}}{H^2(2f_Q+3 Q f_Q)}\Big[E''+18H^3E'+k^2(3H^2-k^2)E\nonumber\\&-9H\varphi'+3(k^2-3H^2)\varphi\Big]=0.
\end{align}
We can see that $\varphi$ is not-dynamical and can be solved for from the first equation. We can then plug the solution into the second equation and corroborate that it is indeed identically satisfied, thus leaving a trivial scalar sector.

The loss of kinetic terms might signal the presence of a strong coupling problem because, as we go arbitrarily close to those maximally symmetric backgrounds, the canonically normalised mode becomes arbitrarily strongly coupled (either to itself or to other modes). It is interesting to notice that these background configurations are in fact the expected solutions in vacuum, since deviations from $\epsilon=0$ will require a non-trivial background profile for the scalar field. For a general FLRW background, however, all three scalar modes propagate.  In other words, in the absence of the matter fields, the disappearance of the scalar modes signals towards a potential strong coupling problem. At a more practical level, these findings suggest the existence of a low strong coupling scale, but it remains to compute it explicitly to see if it leaves room for non-trivial phenomenologies since, at the strong coupling scale, the perturbation theory breaks down and we cannot make any reliable statement. Nevertheless, the situation is better than in $f(T)$ theories since there the strong coupling problem appears for general FLRW backgrounds, i.e., regardless the presence or type of matter fields. Here, we encounter this problem only for (anti-)de Sitter/Minkowski backgrounds and the presence of matter fields can crucially help to raise the strong coupling scale.

It is interesting to compare our results for the perturbations in $f(Q)$ theories with what happens in single field inflation. Let us recall that the curvature perturbation\footnote{Let us stress that $\mathcal{R}$ denotes the scalar perturbation and not the Ricci scalar. Since this is standard notation and there is no risk of confusion, we prefer to stick to the general notation within the inflationary literature.} $\mathcal{R}$ for a single field inflationary model features the following quadratic action
\be
\mS_{\mathcal{R}}= M_{\rm Pl}^2\int\diff\tau\diff^3x\frac{a^2\epsilon}{c_s^2}\left[\dot{\mR}^2-c_s^2\vert \nabla\mR\vert^2\right]\,,
\ee
where $c_s^2$ stands for the propagation speed of the scalar perturbations. In this action we encounter something similar to our $f(Q)$ theories in that the limit $\epsilon\rightarrow 0$ seems to lead to a strong coupling issue because the scalar perturbation loses its kinetic term. This is in fact true. However, in the strict limit $\epsilon=0$, we recover that the matter sector is a cosmological constant and the disappearance of the scalar mode is associated to the fact that a cosmological constant does not fluctuate, so in this particular case there is no onus. The case of the $f(Q)$ theories is more problematic because we have the disappearance of two scalar modes. The disappearance of one of the modes could have been expected on the grounds that the matter sector reduces to a non-fluctuating cosmological constant for $\epsilon=0$. However, we encounter that a second scalar mode also disappears and this mode is expected to be part of the dynamical degrees of freedom of the $f(Q)$ theories so its disappearance in fact signals that it becomes strongly coupled. We will return to this point below where we will relate these evanescent modes around maximally symmetric backgrounds with the appearance of a gauge symmetry. Let us nevertheless finalise this brief comparison with usual inflationary perturbations by saying that very small values of the slow roll parameter can in fact incur strong coupling issues. As a matter of fact, going through a phase where the slow roll parameter decreases for a few e-folds after the CMB scales have exited the horizon and before the end of inflation originates a peak in the power spectrum (that could eventually form primordial black holes). If $\epsilon$ becomes sufficiently small, the peak in the power spectrum can be so large that perturbation theory breaks down.

\subsection{Field equations}\label{Sec:equations}
In the following we will consider a perfect fluid for the matter field instead of a $K$-essence field and explicitly show the dependence of background variables in the perturbations equations of motion. The perturbed line element will be explicitly given in (\ref{conformal}). It is sometimes convenient to introduce the Bardeen potential $\psi$ with the purely temporal transformation
\be
\psi \rightarrow \psi + \mH\zeta^0\,, \label{cgt5}
\ee
by combining the trace of the shear perturbation $E$ with the $\varphi$ on the spatial diagonal as  
\be
\psi = \varphi + \frac{1}{3}\delta^{ij}E_{,ij}\,,
\ee
so that
\begin{align}
\frac{\diff s^2}{a^2(\tau)}  &=  -\lp 1+ 2\phi\rp\diff \tau^2 + 2\lp B_{,i} + B_i\rp \diff\tau\diff x^i \nonumber \\
&+\Big[ \lp 1-2\psi\rp\delta_{ij}+ 2E_{,ij} + 2E_{(i,j)} + 2h_{ij}\Big]\diff x^i\diff x^j\,. \label{conformal}
\end{align}

The matter energy momentum tensor we parameterise in the fluid form,
\ba
T^0{}_0 &= & -\rho\lp1+\delta\rp\,, \\
T^0{}_i &= & -\lp\rho+p\rp \lp \partial_iv + V_i\rp\,, \\
T^i{}_0 &= & \lp\rho+p\rp\lb  \partial^i(v-B) + V^i-B^i\rb\,, \\
T^i{}_j & = & \lp p+\delta p\rp\delta^i_j + \partial^i\partial_j\pi-{\footnotesize{\frac{1}{3}}}\delta^i_j\nabla^2\pi + \Pi^i{}_j\,.
\ea
Here $\rho$ and $p$ are the background energy density and pressure. Denoting $w=p/\rho$, $c_s^2=p'/\rho'$, the continuity and Euler equations are
\ba
\delta'  =   \lp 1+w\rp\lp-k^2v-k^2B+3\varphi'\rp + 3\mH\lp w\rho-\frac{\delta p}{\rho}\rp, \,\,\,\,\label{eqc0} \\
v'   =   -\mH\lp 1-c_s^2\rp v+ \frac{\delta p}{\rho + p} - \frac{2}{3}\frac{w}{1+w}k^2\pi + \phi\,. \qquad \label{eqci}
\ea 
Similarly, the energy constraint can be expressed in a compact form as
\ba
-a^2\delta\rho &=& 6\lp f_{Q} + 12a^{-2}\mH^2 f_{QQ}\rp\mH\lp \mH\phi + \varphi'  \rp + 2f_{Q} k^2\psi  \nonumber\\
&&- 2\lb f_{Q} + 3a^{-2}f_{QQ}\lp\mH'+\mH^2\rp\rb\mH k^2B\,. \label{eq00}
\ea
The velocity propagation can be withdrawn from
\ba
&&\frac{1}{2}a^2\lp\rho+p\rp v = \lb f_{Q} + 3a^{-2}f_{QQ}\lp \mH'+\mH^2\rp\rb\mH \phi   \nonumber\\
&&+ 6a^{-2}f_{QQ}\mH^2\varphi' - 9a^{-2}f_{QQ}\lp\mH'-\mH^2\rp\mH\varphi  \nonumber\\
&&+  f_{Q}\psi'-a^{-2}f_{QQ}\mH^2 k^2B\,. \label{eq0i}
\ea
The pressure equation is a little bit more cumbersome but can be expressed as
\begin{widetext}
\ba
\frac{1}{2}a^2\delta p & = & \lp f_{Q} + 12a^{-2}f_{QQ} \mH^2\rp \lp \mH\phi' + \varphi''\rp + 
\lb f_{Q}\lp \mH'+2\mH^2-\frac{1}{3}k^2\rp + 12a^{-2}f_{QQ}\mH^2\lp 4\mH'-\mH^2\rp + 12 a^{-2}\frac{\diff f_{QQ}}{\diff \tau}\mH^3\rb\phi \nn \\          
& + & 2\lb f_{Q} + 6a^{-2}f_{QQ}\lp3\mH'-\mH^2\rp + 6a^{-2}\frac{\diff f_{QQ}}{\diff \tau}\mH\rb\mH\varphi'  + \frac{1}{3} f_{Q}k^2\psi \nn \\
& - &  \frac{1}{3}\lp f_{Q} + 6a^{-2}f_{QQ}\mH^2\rp k^2 B'
- \frac{1}{3}\lb 2f_{Q}+3a^{-2}f_{QQ}\lp 5\mH-\mH^2\rp + 6a^{-2}\frac{\diff f_{QQ}}{\diff \tau}\mH\rb \mH k^2B\,. \label{eqii}
\ea
\end{widetext}
In the presence of anisotropic stress in $T^i{}_j $, we can similarly follow the shear propagation via the equation
\ba
&&p a^2\pi =
  - 2\Big[ f_{Q} + 6a^{-2}f_{QQ}\lp\mH'-\mH^2\rp\Big]\mH B \nonumber\\
&&+ 2\lb f_{Q}  + 6a^{-2}f_{QQ}\lp\mH'-\mH^2\rp\rb\mH E'  \nonumber\\
&&f_{Q}\lp\psi-\phi\rp+ f_{Q}  E'' - f_{Q}B' \,. \label{eqij}
\ea
Last but not least, we also have the contributions coming from the connection field equations. The two independent connection field equations from (\ref{cfe}) are 
\ba
&-& f_{QQ} \mH \lb 2\mH \varphi' + \lp \mH' + \mH^2\rp\phi + \lp \mH'-\mH^2\rp\lp \psi - B'\rp \rb \nn \\ 
&-& \Big[  f_{QQ} \lp \mH'{}^2 +\mH \mH'' - 3\mH^2 \mH' -\frac{1}{3}\mH^2 k^2\rp \nn \\ &+& 
 \frac{\diff f_{QQ}}{\diff \tau} (\mH'-\mH^2)\mH\Big]B = 0\,, \label{eqc0}
\ea
and
\begin{widetext}
\ba 
&- &f_{QQ} \lp \mH'-3\mH^2\rp\mH\phi'  -  
\lb f_{QQ}\lp \mH''\mH + \mH'{}^2 - 9\mH' \mH^2\rp - \frac{\diff f_{QQ}}{\diff \tau}\lp\mH'-3\mH^2\rp\mH\rb \phi \nn \\
+\quad 2f_{QQ}\mH^2\varphi'' & + & 
 \lb f_{QQ} \lp \mH'+3\mH^2\rp  +  2\frac{\diff f_{QQ}}{\diff \tau}\mH \rb\mH\varphi'  -3   
\lb f_{QQ} \lp\mH'{}^2+\mH''\mH-3\mH'\mH^2\rp  + \frac{\diff f_{QQ}}{\diff \tau}\lp\mH'-\mH^2\rp\mH\rb\varphi \nn \\
- \quad \frac{1}{3}f_{QQ} \mH^2 k^2B' & + & \frac{1}{3}\lb f_{QQ}\lp\mH'-3\mH^2\rp - \frac{\diff f_{QQ}}{\diff \tau} \mH\rb\mH k^2 B =0\,. \label{eqci}
\ea 
\end{widetext}
In terms of the gauge-invariant vector perturbations $\hat{B}^i=(B-E')^i$, the vector equations satisfy
\ba
\lp\rho + p\rp V^i  &=& \frac{1}{2}k^2 \hat{B}^i\,, \\
p\Pi^i  &=&  f_{Q}\lp \hat{B}^i{}'+2\mH \hat{B}^i{}\rp \nonumber\\
&&+ 6a^{-2}f_{QQ}\mH\lp\mH'-\mH^2\rp \hat{B}^i\,. 
\ea
Equipped with these equations of motion of the perturbations we can derive the effects in the cosmological observables, one of the important ones being the change in the growth of structures.

\subsection{Growth of matter perturbations}

We have two extra perturbation variables in the metric compared to the usual case. We also have two extra equations. At small scales, we have from (\ref{eqc0}) 
that\footnote{A similar relation is found for the extra perturbation $\zeta$ in the $f(T)$ models, which however is exact \cite{Golovnev:2018wbh}, whereas the equation (\ref{eqc02}) applies only at the limit $k \gg \mH$.} 
\be \label{eqc02}
k^2 B = 3\mH^{-1}\lb 2\mH\varphi'+\lp\mH'+\mH^2\rp\phi + \lp\mH'-\mH^2\rp\psi\rb\,. 
\ee
Let us consider dust, $w=c_s^2=0$. In the quasi-static limit\footnote{Let us stress that the strict quasi-static limit is not well-defined within these theories because, as we have seen and we will discuss in more depth below, the scalar sector becomes strongly coupled around maximally symmetric backgrounds. Since the quasi-static limit essentially amounts to considering a static background, it is likely that the strong coupling scale is reached before the quasi-static regime. The reason to proceed as we do in this section is to give a comparison with the results in $f(T)$ theories and explicitly show that the perturbations evolve differently even though the background evolutions are formally identical in both families of theories. It is important to notice that the quasi-static limit is even more ill-defined in the $f(T)$ theories due to the more severe strong coupling problems of those theories.} (which is just the small scale approximation), the energy constraint (\ref{eq00}) yields the usual Poisson equation with a modulated Newton's constant,
\be \label{newton}
\psi = -\frac{4\pi G\rho \delta}{k^2f_{Q}}\,,
\ee
where the Newton's constant is explicitly restored ($f_{Q}\rightarrow f_{Q}/8\pi G$). Both the equations of the spatial diagonal (\ref{eqii}) and (\ref{eqij}) both imply at the quasi-static limit that
\be
\phi = \psi\,.
\ee
The connection equation (\ref{eqc02}) then further simplifies to
\be \label{betasol}
B = \frac{6}{k^2}\lp \varphi' + \frac{\mH'}{\mH}\phi\rp\,.
\ee
On the other hand, the continuity equation (\ref{eqc0}) implies for dust that
\be
\delta'' = -k^2\lp v' +B'\rp + 3\varphi''\,,
\ee
which, by using the Euler equation (\ref{eqci}) and then again (\ref{eqc0}) becomes
\be
\delta'' + \mH\delta' + k^2\phi = 3\lp \varphi'' + \mH\varphi'\rp -k^2\lp B'+\mH'B\rp\,.
\ee
Consistently with (\ref{newton}) and (\ref{betasol}), the right hand side of this equation can be neglected at the quasi-static limit. We thus end up with the simple evolution equation for the overdensity $\delta$,
\be
\delta'' + \mH\delta' =  \frac{4\pi G\rho}{f_{Q}} \delta\,,
\ee
where the modification with respect to the standard equation is that the effective gravitational constant is modulated by the time-dependent background function $f_{Q}$ such that $G\rightarrow G/f_{Q}$.
This is essentially the same equation that governs the growth of structures in the quasi-static limit of $f(T)$ cosmology \cite{Golovnev:2018wbh}.

\section{Gauge transformations of the perturbations}
\label{gauge}

In order to appreciate better the involved symmetries, we will consider gauge transformations of the cosmological perturbations in this section. We shall again consider the generally perturbed flat FLRW line element in equation \eqref{perturbed_metric}. We can derive the gauge transformations of the metric potentials under an infinitesimal coordinate change $x^\mu \rightarrow x^\mu + \zeta^\mu$, whose scalar sector is given by $\zeta^0$ and {$\zeta^i=\delta^{ij}\partial_j\zeta$} and vector sector by a transverse $\zetah^i$. The result for the vectors is
\ba
B^i & \rightarrow & B^i - \zetah^i{}'\,,\nonumber \\
E^i & \rightarrow & E^i - \zetah^i{}\,,
\ea
and for the scalars
\ba
\phi & \rightarrow & \phi -{\zeta^0}' - \mH \zeta^0\,, \label{cgt1} \nonumber\\
B & \rightarrow & B - \zeta' + \zeta^0\,, \nonumber\\
\varphi& \rightarrow & \varphi - \frac{1}{3}k^2\zeta + \mH\zeta^0\,, \nonumber\\
E & \rightarrow & E - \zeta\,.
\label{gaugetransf}
\ea
As we have explained in several occasions above, the use of the coincident gauge exhausted the freedom allowed by Diff-invariance. This is crucially different from other Diff-invariant theories (as e.g. GR and many other extensions based on its curvature formulation), but it is somewhat similar to what happens for massive gravity theories where the mass term breaks Diffs invariance and it is not legitimate to fix any of the gravitational potentials (some of which remain non-dynamical like in our case). The analogy with massive gravity also extends to the Diffs-restored versions of the theories. In massive gravity it is possible to resort to the Stueckelberg trick and restore Diff-invariance by the introduction of a set of four compensating fields with appropriate transformation properties (under both Lorentz and diffeomorphisms transformations), while in symmetric teleparallel theories the connection naturally acquires the form of an inertial connection given in terms of compensating Stueckelberg fields, the $\xi's$, and the coincident gauge is nothing but the corresponding unitary gauge. It will be instructive to explicitly see how the form of the inertial connection indeed restores the scalar character of the non-metricity scalar in the cosmological framework. This will also permit us to gain a better understanding of the disappearance of scalar modes around maximally symmetric backgrounds obtained above as a consequence of the emergence of a constrained Diff symmetry for such backgrounds.

The non-metricity scalar in the coincident gauge for our perturbed metric at first order is\footnote{We will restore the notation of denoting a quantity in the coincident gauge with a ring over for clarity.}
\be \label{Qscalar}
a^2 \Qr = 6\mH^2 - 2\mH\lb 6\lp \mH\phi+\varphi'\rp - k^2B\rb\,,
\ee
Notice that vector perturbations do not contribute at this order, so we will not consider them for now, but we will come back to them later when discussing the invariance of the field equations.

The transformation of the non-metricity perturbation in the coincident gauge  can then be easily computed as\footnote{In this section we will use $\Delta_\zeta$ to denote the variation under a Diff, while $\delta$ will be used to denote the first order perturbation of a given quantity.}
\ba \label{changeQ}
\Delta_\zeta \delta\Qr=  \frac{2\mH}{a^{2}}\Big[ 6\lp\mH^2-\mH'\rp\zeta^0 +k^2\lp\zeta^0+\zeta'\rp \Big].
\ea
As expected, $\Qr$ does not transform as a scalar but it features an additional term. It is interesting however, that the failure to transform as a scalar is
\be
\Delta_\zeta\delta\Qr+\mL_\zeta\Qr=2\mH\frac{k^2}{a^{2}}\lp\zeta^0+\zeta'\rp
\ee
which shows three remarkable properties. Firstly, $\Qr$ does transform as a scalar for {\it large gauge transformations} with zero momentum and this can have interesting consequences for adiabatic modes. Secondly, non-zero momentum gauge transformations still realise a restricted Diff symmetry provided the gauge parameters satisfy $\zeta^0=-\zeta'$.  Notice however that this does not mean the existence of a gauge symmetry in the theory. Finally, the Minkowski background with $\mH=0$ also makes $\delta\Qr$ transform as a scalar. This already suggests that Minkowski is a special background as we will confirm in the following. In order to elucidate more clearly the role of the gauge transformations on the dynamics, let us see how the equations for the perturbations behave under Diffs. We do not need the specific form of the equations, but only how they change under a Diffs transformation. As above, we will use a scalar field as a proxy for the matter sector. Since this matter sector is Diff-invariant, its equation of motion will be gauge invariant so we do not need to consider it and we can only focus on the gravitational equations for the perturbations, which correspond to the Diff-breaking sector. These equations change as:
\begin{align}
\Delta_\zeta\mathcal{E}_\varphi=&-2k^2\bar{Q} f_{QQ}(\zeta^0+\zeta')'+\epsilon k^2\mH \bar{Q}f_{QQ}(5\zeta^0+7\zeta')\nonumber\\
&+4k^2\mH\bar{Q}^2f_{QQQ}(\epsilon-1)(\zeta^0+\zeta')\,,\nonumber\\
\Delta_\zeta \mathcal{E}_\phi=&k^2\mH \bar{Q}f_{QQ}\Big[2(\zeta^0+\zeta')+\epsilon(\zeta^0-\zeta')\Big]\,,\nonumber\\
\Delta_\zeta\mathcal{E}_B=&\frac{1}{3}k^2\bar{Q} f_{QQ}\left[3\epsilon \mH \left({\zeta^0}' +k^2\zeta\right)-k^2\left(\zeta^0+\zeta' \right) \right]\,,\nonumber\\
\Delta_\zeta\mathcal{E}_E=&-\frac{4\epsilon}{3} k^4 \mH\bar{Q} f_{QQ} \zeta^0\,.
\label{eq:DeltaEqscalars}
\end{align}
From these equations we can see that, as expected, the STEGR where $f(Q)$ is a linear function of $Q$ the equations are gauge invariant. However, we can also see that background solutions with $\epsilon=0$ also have perturbations equations that are invariant under the restricted Diffs with $\zeta^0=-\zeta'$, which is the same subset of Diffs that make $\delta\Qr$ transform as a scalar. This feature also allows to explain the disappearance of the scalar modes discussed above around maximally symmetric backgrounds. We can now understand that these backgrounds enjoy an additional gauge symmetry so that one scalar perturbation becomes a gauge mode. However, since this symmetry is not in the full theory, we can conclude that it is just an accidental gauge symmetry of those backgrounds. The accidental nature of this symmetry is in fact corroborated by the fact that more general cosmological backgrounds do not have it, which can be easily understood by noticing that there is no relation between $\zeta^0$ and $\zeta$ that make the RHS in (\ref{eq:DeltaEqscalars}) vanish for an arbitrary background. This is precisely the origin of the aforementioned strong coupling problem of the scalar perturbations around maximally symmetric backgrounds. On the other hand, the fact that Minkowski exhibits this enhanced symmetry also explains that the very sub-horizon modes realise the symmetry up to corrections suppressed by $\mH^2/k^2$.

Let us now turn to the vector sector whose equations we gave in (\ref{eq:vecpert}). From there, it is obvious to see that they do retain the gauge invariance of GR for an arbitrary purely vector Diff. Whether this symmetry is maintained at higher order is not clear. If they do, the non-dynamical nature of the vector modes would extend to the full non-linear order. If a completion of this linear symmetry is not present, then vector modes would exhibit strong coupling problems more severe than those of the scalar perturbations.

To end this section, it is instructive to see how the above gauge transformations relate to the choice of connection.
At the background order the non-metricity can be written as
\be \label{bQ}
Q_{\alpha\mu\nu} = 2\mH \delta^0_\alpha g_{\mu\nu}\,, \quad  Q_\alpha = 8\mH\delta^0_\alpha\,, \quad \tilde{Q}_\alpha = 2\mH\delta^0_\alpha\,,
\ee
and for its conjugate we obtain 
\be \label{bconjugate}
P_{\alpha\mu\nu} = \mH\lb \delta^0_\alpha\delta^i_\mu\delta^j_\nu - \delta_\alpha^i\delta^j_{(\mu}\delta^0_{\nu)}\rb g_{ij}\,.
\ee
If we transform the connection away from the coincident gauge by the same $\zeta^\mu$ as in \eqref{gaugetransf}, up to the linear order we have 
$\delta Q_{\alpha\mu\nu} = -2g_{\beta(\mu}\partial_{\nu)}\partial_\alpha\zeta^\beta$ and then, from $Q=-P_{\alpha\mu\nu}Q^{\alpha\mu\nu}$ we obtain, by using (\ref{bconjugate}), that $ \delta Q = -2\mH\lp \zeta^0{}_{,ij}+ \zeta_{i,(j0)}\rp g^{ij}$. Thus the variation of $Q$ due to the change of the connection is given by 
\be 
\delta  Q = 2a^{-2}\mH k^2\lp \zeta^0 + \zeta'\rp,
\ee  
which precisely cancels the non-scalar variation in  (\ref{changeQ}) and we obtain $\Delta_\zeta \delta Q=-\mL_\zeta Q$, i.e., it does transform as a scalar.

Any other contraction of the non-metricity, besides $Q$, will be a scalar too. Let us check for example the term
\be
a^2Q_1 \equiv a^2 Q_{\alpha\mu\nu}Q^{\alpha\mu\nu} = -16\mH^2  + 8\mH\lp 4\mH\phi-\phi'+3\varphi'\rp\,,
\ee
which under (\ref{gaugetransf}) transforms in the coincident gauge as
\be \label{transA}
Q_1 \rightarrow Q_1  + 8\mH a^{-2}\lb {\zeta^0}{}''+4 \lp \mH'-\mH^2\rp\zeta^0 - k^2\zeta' \rb
\ee
In this case we immediately see from (\ref{bQ}) that the accompanying change of the connection away from the coincidence gauge gives
\be
\delta Q_1 = 2Q^{\alpha\mu\nu}2g_{\beta(\mu}\partial_{\nu)}\partial_\alpha\zeta^\beta =-8\mH a^{-2}\lp \zeta^0{}'' -  k^2\zeta'\rp, 
\ee
so the total change given by the sum of the above two contributions (i.e. the change in the coincident gauge plus the change of the connection) is
\be
\Delta_\zeta Q_1=\frac{32}{a^2}\mH(\mH'-\mH^2)\zeta^0
\ee
which is nothing but $\Delta_\zeta Q_1=-\zeta_0 \bar{Q}_1'=-\mL_\zeta \bar{Q}_1$ as it should for a scalar. It is straightforward to show that any other contraction also transforms as a scalar when the change in the connection is properly taken into account.

\section{Conclusions}
\label{conclusions}
In this work we have studied the cosmological implications of the new type of modified gravity theories that originate from the equivalent formulation of GR based on non-metricity, namely: The Symmetric Teleparallel Equivalent of GR. Models of general functions of the Ricci or torsion scalar have already been extensively studied in the literature. The cosmological realisation of $f(R)$ theories tends to force to remain close to GR, whereas models based on $f(T)$ suffer from strong coupling problems on general FLRW backgrounds. The third equivalent formulation of GR by means of the $Q$-scalar motivates novel ways of modifying gravity, one of such examples being the $f(Q)$ theories. Even if these models are indistinguishable from $f(T)$ theories at the background level, crucial differences arise at the level of cosmological perturbations, which has been the aim of this work. Specially, we have shown that the strong coupling problems possibly encountered in $f(T)$ theories are absent in $f(Q)$ models on general FLRW backgrounds. However, they do appear on maximally symmetric backgrounds such as Minkowski and de Sitter. Although these highly symmetric backgrounds might suffer from such strong coupling problems, they are less severe than the ones exhibited by $f(T)$ theories and, given that less symmetric cosmologies do not suffer from these problems, there can be room for interesting phenomenologies. We have also shown that at the small-scale quasi-static limit the predictions of the $f(Q)$ and the $f(T)$ models coincide, but at larger scales the $f(Q)$ models generically propagate two scalar degrees of freedom that are absent in the case of $f(T)$. These two degrees of freedom are the ones that disappear around maximally symmetric backgrounds and, thus, cause the discussed strong coupling problem. 

Since the $f(Q)$ theories in the coincident gauge do not have the usual gauge invariance of cosmological perturbations, we have performed a detailed analysis of the behaviour of the equations under Diffs transformation. Since the theory is no longer Diff-invariant, the equations do change under a Diff. Remarkably, we have found that maximally symmetric backgrounds retain a gauge symmetry given by a restricted Diff. These findings allowed us to give a better understanding on the disappearance of degrees of freedom around these backgrounds as a consequence of the appearance of a residual gauge symmetry, which then roots the strong coupling problems.

There is an important caveat that has to be taken into account before drawing conclusions about the viability of $f(Q)$ cosmology. Because in this framework the connection is an independent fundamental degree of freedom besides the
metric, the space of solutions is richer than in purely metric gravity. In particular, there exists physically inequivalent cosmological solutions that respect the FLRW symmetry at the background level, but whose perturbations
may behave differently\footnote{An analogous issue has been recently pointed out concerning the $f(T)$ and other modified teleparallel gravity models. There the number and the nature of the degrees of freedom can depend
on the Lorentz frame wherein one chooses to introduce the fluctuations \cite{Golovnev:2018wbh,Koivisto:2018loq}.}. In the case $f(Q)=Q$ the dynamics are frame-independent, but once $f_{QQ}(Q) \neq 0$ an ambiguity arises.
Recently, we have put forward the conjecture that in the {\it canonical frame} the energy-momentum of the metric field is vanishing \cite{Jimenez:2019yyx}. This is a covariant criterion that eliminates the ambiguity in the 
predictions of the modified models. The question arises whether the strong coupling issue on Minkowski and de Sitter backgrounds could be avoided by the more judicious choice of the background solution for the metric and the connection. This calls for reconsideration of the $f(Q)$ cosmology in the canonical frame, a task we plan to undertake in the near future. 
\\
\acknowledgments  JBJ acknowledges support from the  {\it Atracci\'on del Talento Cient\'ifico en Salamanca} programme and the MINECO's projects FIS2014-52837-P and FIS2016-78859-P (AEI/FEDER). LH is supported by funding from the European Research Council (ERC) under the European Unions Horizon 2020 research and innovation programme grant agreement No 801781 and by the Swiss National Science Foundation grant 179740.
TK acknowledges support from the Estonian Research Council grant PRG356 ``Gauge Gravity'' and the European Regional
Development Fund through the Center of Excellence TK133 ``The Dark Side of the Universe''. This research is based upon work from COST Action CA15117 CANTATA, supported by the European Cooperation in Science and Technology.


\appendix

\section{Upper bound on the dynamical degrees of freedom}\label{ADM}
In this appendix we will elaborate on the argument that establishes a maximum of six propagating modes in the full $f(Q)$ theories. For that we will resort to the ADM analysis that is based on the following decomposition of the metric
\be
\diff s^{2}=-\left(\mN^{2}-\mN_{i} \mN^{i}\right) \diff t^{2}+2 \mN_{i} \diff x^{i} \diff t+\gamma_{i j} \diff x^{i} \diff x^{j}
\ee
where $\mN$ is the lapse function, $\mN_i$ is the shift vector and $\gamma_{ij}$ is the spatial metric, which is in turn used to lower and raise spatial indices. It will be convenient to work in the coincident gauge where the full connection vanishes and the non-metricity scalar can be written in terms of the Christoffel symbols of the metric as follows:
\be
Q=g^{\mu\nu}\Big(\left\{^{\phantom{i} \alpha}_{\alpha\beta}\right\}\left\{^{\phantom{i} \beta}_{\nu\mu}\right\} -\left\{^{\phantom{i} \alpha}_{\nu\beta}\right\} \left\{^{\phantom{i} \beta}_{\alpha\mu}\right\} \Big).
\label{eq:QCG}
\ee
The crucial point now is to realise that (\ref{eq:QCG}) is precisely the so-called $\Gamma\Gamma$ part of the metric Ricci scalar as can be easily identified from the general expression
\be
R=g^{\mu\nu}\Big(\partial_\alpha \Gamma^\alpha{}_{\nu\mu}-\partial_\nu \Gamma^\alpha{}_{\alpha\mu}+\Gamma^\alpha{}_{\alpha\beta}\Gamma^\beta{}_{\nu\mu}-\Gamma^\alpha{}_{\nu\beta}\Gamma^\beta{}_{\alpha\mu}\Big)
\ee
so we can utilise the usual results from the ADM analysis of GR by simply dropping the terms containing second derivatives that originate from the piece $\partial\Gamma$ in the Ricci scalar. To be explicit, let us take the general ADM decomposition of the Ricci scalar
\begin{align}
R(g)=&\,^{(3)}R+K_{ij} K^{ij}+K^2\nonumber\\
&-\frac{2}{\mN}\Big(\dot{K}^i{}_i-\mN^i\,^{(3)}\nabla_i K+\,^{(3)}\nabla^2\mN\big)
\end{align}
with $K_{ij}=-\frac12\mN^{-1}\Big(\dot{\gamma}_{ij}-2\,^{(3)}\nabla_{(i}\mN_{j)}\Big)$ the extrinsic curvature, $K$ its trace and $\,^{(3)}R$ the Ricci scalar of $\gamma_{ij}$. Thus, we can straightforwardly obtain the ADM decomposition of $Q$ as
\be
Q=K_{ij} K^{ij}+K^2+\gamma^{kl}\Big(\Gamma^i{}_{ij}\Gamma^j{}_{lk}-\Gamma^i{}_{lj}\Gamma^j{}_{ik}\Big).
\ee
Since the only time derivatives in $Q$ come from $\dot{\gamma}_{ij}$ in $K_{ij}$, we can directly conclude that $\mN$ and $\mN_i$ will be non-dynamical fields in the full $f(Q)$ theories and, as a consequence, these theories can only have up to six dynamical modes corresponding to the six components of $\gamma_{ij}$. To unveil the full dynamical content of the theories, a complete Hamiltonian analysis would be necessary and the number of propagating modes will depend on the specific function form of $f$. For instance, for constant $f_Q$ we will recover that the shift and the lapse enforce the first class constraints associated to diffeomorphisms. Moreover, the Poisson algebra will also exhibit singular points for some background configurations that will feature accidental gauge symmetries as it occurs for the maximally symmetric configurations considered in this work. A similar feature appears in $f(T)$ theories \cite{Ferraro:2014owa}.

\section{Full Hessian for the scalar sector}

For completeness, in this appendix we will give the full expression for the Hessian of the scalar perturbations. Having a positive definite Hessian will be required to guarantee the absence of ghost-like degrees of freedom in the scalar sector and this will impose additional stability constraints. The expression for the Hessian after integrating out the non-dynamical scalar modes is given by
\be \label{fullhessian}
H=
\frac{1}{g}\left(
\begin{array}{ccc}
 a &b   &c   \\
  b & d  & e   \\
  c & e  & f-g
\end{array}
\right)\,,
\ee
where
\be
g = 12f_Q^2+2\lp 2-5\epsilon\rp f_QQf_{QQ}- \epsilon\lp 8-3\epsilon\rp Qf_{QQ}^2\,, \nn 
\ee
and
\ba
a & = & -6\epsilon\lp f_{Q} + 2Qf_{QQ}\rp\lp4f_Q^2+6f_Q Qf_{QQ}-3\epsilon Qf_{QQ}^2\rp\,,\nn \\
b & = & 4\epsilon k^2 f_Q\lp 2f_Q-QF_{QQ}\rp\lp f_Q+2QF_{QQ}\rp\,, \nn \\ 
c & = & -6\sqrt{ 2\epsilon\lp f_Q+2Qf_{QQ}\rp}\lb 2f_Q^2+\lp 3-\epsilon\rp f_QQf_{QQ} - \epsilon Qf_Q^2\rb\,, \nn \\
d & = & \frac{2}{3}k^4 f_Q\lb 2\lp 3+\epsilon\rp f_QQf_{QQ} + \lp 8-3\epsilon\rp\epsilon QF_{QQ}^2 -4\epsilon f_Q^2 \rb\,,\nn\\
e & = & 2k^2f_Q\sqrt{2\epsilon\lp f_Q+2Qf_{QQ}\rp}\lb 2f_Q+\lp 2-\epsilon\rp Qf_{QQ}\rb\,,\nn\\
f & = & 2Q f_{QQ}\lp f_Q+2 Q f_{QQ}\rp\,. \nn
\ea
There is no simple statement about the stability that can be extracted from this, unless some special limits and/or functions are taken. The determinant is given by (\ref{hessiandet}).

\section{Bianchi I universe}

In this appendix we will briefly discuss the cosmology of the $f(Q)$ theories for anisotropic universes in order to explicitly show that universes close to isotropic solutions will isotropize as in more conventional cases. This will show that the pathological behaviour of the perturbations is not captured by simple deformations of the FLRW universes. The Bianchi I metric is described by the line element
\be
\diff s^2=-N^2(t)\diff t^2+a_1^2(t)\diff x^2+a_2^2(t)\diff y^2+a_3^2(t)\diff z
\ee
with $a_i(t)$ the scale factors along the three spatial directions. We will introduced the isotropic scale factor defined as $a^3\equiv a_1 a_2a_3$ and the corresponding expansion rates
\be
H_i=\frac{\dot{a_i}}{a_i}, \quad H=\frac{\dot{a}}{a}.
\ee
The action in the corresponding mini-superspace is given by
\be
\mS=-\frac12\int \diff t\,N a^3f(Q)
\label{actionBianchi}
\ee
where the non-metricity scalar reads
\be
Q=2\frac{H_1(H_2+H_3)+H_2H_3}{N^2}.
\ee
It is easy to see that the Bianchi I Ansatz retains the time-reparameterisation invariance so we will set the lapse $N=1$ after having obtained the field equations. The full set of equations can be easily obtained from (\ref{actionBianchi}). However, since we are only interested in studying perturbations around the isotropic case, we will only need the combinations
\begin{eqnarray}
\frac{\delta \mS}{\delta a_1}-\frac{\delta \mS}{\delta a_3}=0,\quad
\frac{\delta \mS}{\delta a_2}-\frac{\delta \mS}{\delta a_3}=0
\label{eqanisotropy}
\end{eqnarray}
that describe the evolution of the anisotropy. Notice that, provided the matter sector does not contain any anisotropic stresses, these equations are not sourced. Furthermore, it will be sufficient for us to consider small perturbations so we will parametrise the small deformation of the FLRW metric as follows
\begin{eqnarray}
H_1&=&H+2 \delta_1-\delta_2\\
H_2&=&H-\delta_1+2\delta_2\\
H_3&=&H-\delta_1-\delta_2
\end{eqnarray}
where $\delta_1$ and $\delta_2$ describe the two independent anisotropies of the Bianchi universe and the parameterisation has been chosen for convenience and to guarantee that $3H=H_1+H_2+H_3$. When plugged into (\ref{eqanisotropy}) and expanded to first order in $\delta_{1,2}$, we obtain the equations
\be
\dot{\delta}_i+3H\left(1+\frac{4\dot{H} f_{QQ}}{f_Q}\right)\delta_i=0,\quad i=1,2.
\ee
This equation shows that an expanding universe will isotropize provided $1+\frac{4\dot{H} f_{QQ}}{f_Q}>0$. We can do better by noticing that the above equation can be rearranged into the form
\be
\frac{\diff}{\diff t}\left(a^3f_Q\delta_i\right)=0
\label{eqdeltas}
\ee
so the stability is entirely determined by the behaviour of $f_Q$. This result is the expected one because a Bianchi I universe describing a small anisotropic deformation of FLRW can be associated to a long wave-length GW and, as we discussed in \ref{tensormodes}, there is no pathological behaviour in that sector as long as $f_Q>0$. In fact, what we have obtained in \ref{eqdeltas} is nothing but the equation for the GWs in the limit $k\rightarrow 0$, as it should be.

\newpage

\bibliography{fQperturbations}

\end{document}